\documentclass[aps,prb,english,twocolumn,showpacs,preprintnumbers,amsmath,amssymb,floatfix,superscriptaddress,longbibliography]{revtex4-1}

\usepackage[T1]{fontenc}
\usepackage[latin9]{inputenc}
\usepackage{graphicx}
\usepackage{amssymb}
\usepackage{babel}
\makeatletter

\usepackage[version=3]{mhchem}

\usepackage{color}

\usepackage{hyperref}
\hypersetup{
     colorlinks = true,
     citecolor  = blue,  
     linkcolor  = magenta 
}

\makeatother
\usepackage{babel}
\makeatother

\begin{document}

\title{A knob to tune the Casimir-Lifshitz force with gapped metals}

\author{M.  Bostr{\"o}m}
  \email{mathias.bostrom@ensemble3.eu}
 \affiliation{Centre of Excellence ENSEMBLE3 Sp. z o. o., Wolczynska Str. 133, 01-919, Warsaw, Poland}

\author{M. R. Khan}
  \email{muhammadrizwan.khan@ensemble3.eu}
\affiliation{Centre of Excellence ENSEMBLE3 Sp. z o. o., Wolczynska Str. 133, 01-919, Warsaw, Poland}

\author{H. R. Gopidi}
\affiliation{Centre of Excellence ENSEMBLE3 Sp. z o. o., Wolczynska Str. 133, 01-919, Warsaw, Poland}

\author{I. Brevik}
\affiliation{Department of Energy and Process Engineering, Norwegian University of Science and Technology, NO-7491 Trondheim, Norway}

\author{Y. Li}
  \affiliation{Department of Physics, Nanchang University, Nanchang 330031, China}
  \affiliation{Institute of Space Science and Technology, Nanchang University, Nanchang 330031, China}

\author{C. Persson}
\affiliation{Centre for Materials Science and Nanotechnology, Department of Physics, University of Oslo, P. O. Box 1048 Blindern, NO-0316 Oslo, Norway}
\affiliation{Department of Materials Science and Engineering, KTH Royal Institute of Technology, SE-100 44 Stockholm, Sweden}

\author{O. I. Malyi}
\email{oleksandr.malyi@ensemble3.eu}
\affiliation{Centre of Excellence ENSEMBLE3 Sp. z o. o., Wolczynska Str. 133, 01-919, Warsaw, Poland}

\date{\today}%

\begin{abstract}
The Casimir-Lifshitz interaction, a long-range force that arises between solids and molecules due to quantum fluctuations in electromagnetic fields, has been widely studied in solid-state physics. The degree of polarization in this interaction is influenced by the dielectric properties of the materials involved, which in turn are determined by factors such as band-to-band transitions, free carrier contributions, phonon contributions, and exciton contributions. Gapped metals, a new class of materials with unique electronic structures, offer the potential to manipulate dielectric properties and, consequently, the Casimir-Lifshitz interaction. In this study, we theoretically investigate the finite temperature Casimir-Lifshitz interaction in La$_3$Te$_4$-based gapped metal systems with varying off-stoichiometry levels. We demonstrate that  off-stoichiometric effects in gapped metals can be used to control the magnitude and, in some cases, even the sign of Casimir-Lifshitz interactions. We predict measurable corrections due to stoichiometry on the predicted Casimir force between a La$_3$Te$_4$ surface and a gold sphere, attached to an atomic force microscopy tip.
\end{abstract}

\maketitle

\section{Introduction}
Solid-state physics textbooks teach us about the Casimir-Lifshitz interaction\,\cite{Milton2001,Bordagbook,NinhamLoNostro2010book,Ser2018}, as a long-range force arising between solids and molecules due to quantum fluctuations in the electromagnetic fields\,\cite{casimir1948-1,Dzya,ParsegianNinham1969}. This interaction causes the solids or molecules to become polarized when they are close to each other, with the degree of polarization directly influenced by the dielectric properties of the materials \,\cite{Rich71,Rich73,Lamo1997,Bost2000,Bord,PhysRevLett.87.259101,Hoye,KimballAMilton2004,Lamoreaux_2005,Mohid,RevModPhys.81.1827,SushNP,PhysRevB.97.125421,Estesodoi:10.1021/acs.jpclett.9b02030,KlimchitskayaMohideenMostepanenko,Nesterenko2022}. This relationship indicates that by manipulating the dielectric properties through external means, it is possible to make an impact on the Casimir-Lifshitz interaction directly. The effects of such manipulations can be directly measured using atomic force microscopy\,\cite{DuckerSendenPashley_AFM_Nature} or even be employed in the development of functional devices.

As electronic structure theory has advanced, it has become clear that a material's dielectric properties are determined by several factors: (i) band-to-band transitions between occupied and unoccupied states, which play a significant role across all frequency ranges; (ii) free carrier contributions, typically found in metallic systems at low frequencies; (iii) the phonon contribution; and (iv) exciton contributions, with the latter two primarily limited to the low-frequency range. This understanding implies that by identifying ways to influence factors (i) through (iv), we can effectively tune the dielectric properties of materials and, consequently, the Casimir-Lifshitz interaction. Therefore, the primary challenge lies in discovering materials with tunable dielectric properties.

\begin{table*}
\centering
\caption{Parametrization of the average dielectric function of continuous media, $\varepsilon(i\xi)$, at
imaginary frequencies for La$_{3-x}$Te$_4$} as calculated with first-principles calculations and a damping coefficient ($\Gamma$) set to 0.2\,eV. { The $\omega_j$ modes are given in $\rm eV$.} The largest difference between fitted and calculated $\varepsilon(i\xi)$ is 0.04$\%$. 

\begin{tabular}{ p{2.1cm} p{2.1cm} p{2.1cm} p{2.1cm}  p{2.1cm} p{2.1cm} }
\hline
\hline
{modes ($\omega_j$)}  & \multicolumn{4}{c}{ $C_j$ for different La$_{3-x}$Te$_4$ compounds}  \\
\hline
   & La$_3$Te$_4$           & La$_{2.92}$Te$_4$   & La$_{2.83}$Te$_4$       & La$_{2.75}$Te$_4$       & La$_{2.67}$Te$_4$ \\
\hline
 0.0203        & 5.6047    &  24.2875       &  0.0    &   0.0  & 0.0  \\

\hline
 0.0362       & 15.3667   &  55.9154        &  0.0   &   0.0 &  0.0  \\

\hline
 0.0694        &27.1114     &  98.6417       &  1.2411   &   0.4796 & 0.0008 \\

\hline
 0.1325       & 35.8341    &  72.487       &  42.8899 &   19.6292  & 0.0  \\

\hline
 0.2085      &93.1702     & 47.7692       &   42.1523 &   25.6095 & 0.0   \\

\hline

0.4297        &6.5149      & 7.3256       &   5.479 &   2.4194 & 0.0095  \\

\hline

 0.8328       & 2.9159     &  1.447       &   0.7619 &   0.4956 &  0.0421  \\

\hline

 1.9083    &   2.4749    &  2.9666             & 3.022  &    3.3971 &    3.1313  \\

\hline
 3.2388      & 4.6107      &  4.3332      &   4.464 &   4.5286 & 4.8426   \\

\hline

 5.2955      & 2.7231     & 2.8949       &   2.7872 &   2.6545 & 2.4639   \\

\hline

 8.9753       & 0.7087      &  0.5961     &   0.6287 &   0.6727 & 0.7464   \\

\hline

 18.2815  & 0.1959   &  0.2782   &   0.2486 &   0.205 & 0.1581   \\

\hline
 23.0355  & 0.0695   &  0.0069   &   0.022 &   0.0474 & 0.0716  \\
\hline
42.0922  & 0.0   &  0.0057   &   0.0042 &   0.0018 & 0.0   \\
\hline



\hline
\hline
\end{tabular}
\end{table*}

Recently, gapped metals have emerged as a new class of materials possessing unique electronic structures \cite{malyi2019spontaneous,  zhang2015intrinsic, ricci2020gapped, Zhang_etal_NatMat2016, electrideMatsuishietal2003, electride_Mayetal_2008_PhysRevB.78.125205}. These compounds set themselves apart from both metals and insulators, as they possess a Fermi level within the conduction (or valence) band, resulting in a high intrinsic concentration of free carriers with an internal band gap between their primary band edges. {Such  materials have  attracted significant attention in connection with transparent conductors\,\cite{Zhang_etal_NatMat2016}, thermoelectrics\,\cite{electride_Mayetal_2008_PhysRevB.78.125205}, and electrides\,\cite{electrideMatsuishietal2003}. What makes these materials special is that they can develop  off-stoichiometry (within the same parental structure) due to the decay of conducting electrons (holes) to acceptor (donor) states (Fig. 1a,b) formed by intrinsic defect formation \cite{malyi2019spontaneous, doi:10.1021/acs.jpclett.2c03701, malyi2020false, malyi2020realization, zunger2021understanding}.
This is different from the traditional metals and insulators, where defect formation is usually limited to high temperatures and is primarily driven by an increase in configurational entropy \cite{zunger2021understanding}. This situation is unique because different off-stoichiometry levels for gapped metals can be achieved, each exhibiting distinct dielectric properties. For example, La$_3$Te$_4$ - an n-type gapped metal (i.e., one with its Fermi level in the primary conduction band) - can be experimentally synthesized \cite{ramsey1965phase, delaire2009phonon, li2022enhanced} (simply by change synthesis conditions) across a range of phases from La$_3$Te$_4$ to La$_{2.66}$Te$_4$ with their properties tunable from metallic to insulating \cite{doi:10.1021/acs.jpclett.2c03701} as schematically shown in Fig. 1c.} Motivated by the above,   we study here theoretically the finite temperature Casimir-Lifshitz interaction  between different systems involving La$_3$Te$_4$-based gapped metals, demonstrating how off-stoichiometry can be used as a knob to tune such long-range interaction.
\begin{figure*}
  \centering
  \includegraphics[width=0.7\textwidth]{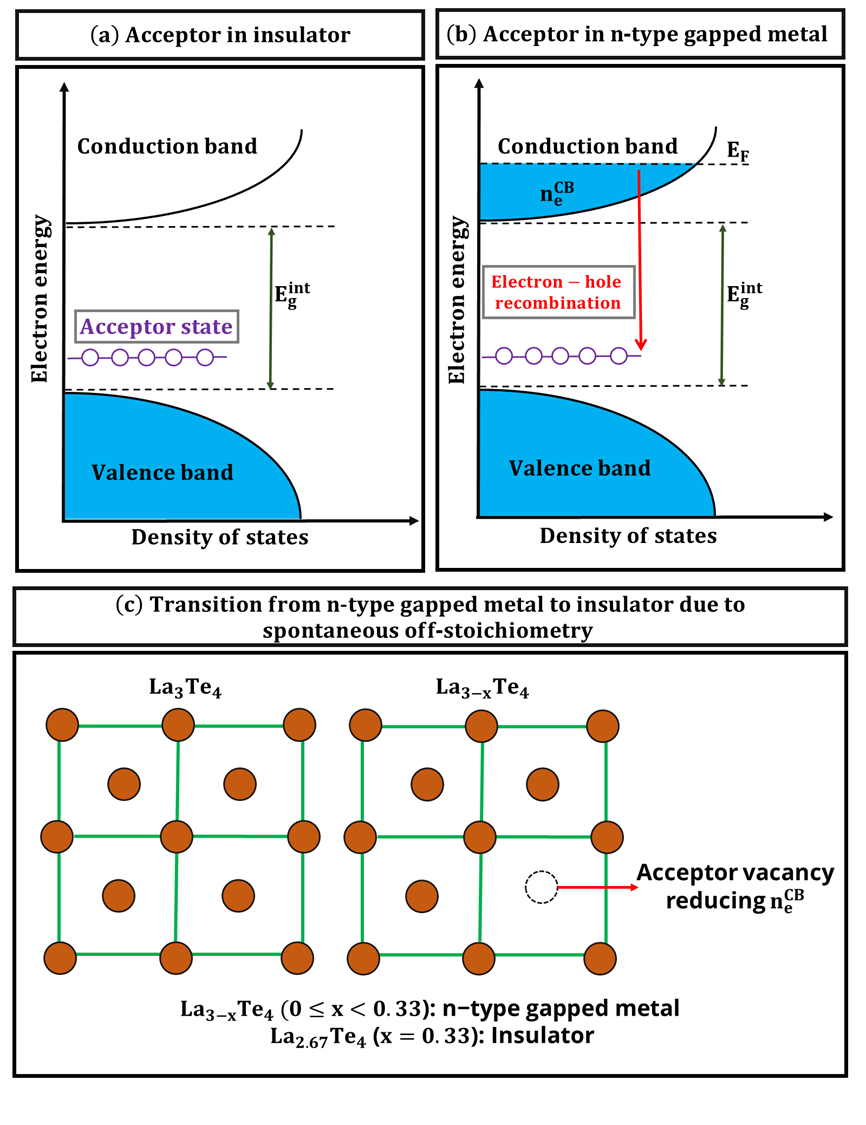}
  \caption{\label{Schematics} (Color online) Origin of off-stoichiometry in La$_3$Te$_4$ compound. Formation of acceptor vacancy in (a) insulator and (b) n-type gapped metal. Here, E$_F$ and E$_g^{int}$ correspond to the Fermi level and internal band gap between the principal valence band maximum and conduction band minimum. (c) Schematic crystal structure for La-Te system depicting metallic state with free electrons in conduction bands and insulator state of La$_{3-x}$Te$_4$ phase due to presence of acceptor vacancy.}
\end{figure*}

\section{Methods}
\subsection{Computational details and dielectric functions}
To compute the dielectric properties of gapped metals, we perform first-principles calculations using the Perdew-Burke-Ernzerhof exchange-correlation (PBE XC) functional \cite{perdew1996generalized} within the VASP framework\,\cite{vasp1,vasp3,vasp4,vasp1999}. Our analysis focuses on five distinct La$_{3-x}$Te$_4$ compounds, previously identified in our earlier work\,\cite{doi:10.1021/acs.jpclett.2c03701}. For each system, we calculate the dielectric properties, considering only the Drude contribution and interband transitions. We employ $\Gamma$-centered Monkhorst-Pack k-grids \cite{monkhorst1976special} with 20,000 points per reciprocal atom for the calculations of direct band transitions and plasma frequencies and introduce a subtle Lorentzian broadening of 0.01 eV in the Kramers-Kronig transformation\cite{landau2013statistical}. To include the Drude term in the optical properties, we utilize the kram code\,\cite{blaha2020wien2k, blaha1990full, blaha2001wien2k}, setting the damping coefficient ($\Gamma$) to 0.2\,eV, and additionally investigate the influence of the $\Gamma$ parameter on our findings. The other details on computational parameters can be found elsewhere\,\cite{doi:10.1021/acs.jpclett.2c03701, khan2023spontaneous}.  {Although we are unable to directly compare our calculated dielectric functions with corresponding experimental measurements (no data are available), it is important to highlight that the used methods can describe sufficiently well the experimental trends related to transparency and coloring across a range of gapped metals. \cite{malyi2019spontaneous,khan2023optical}}

\begin{figure*}
 \includegraphics[width=0.8\textwidth,height=6cm]{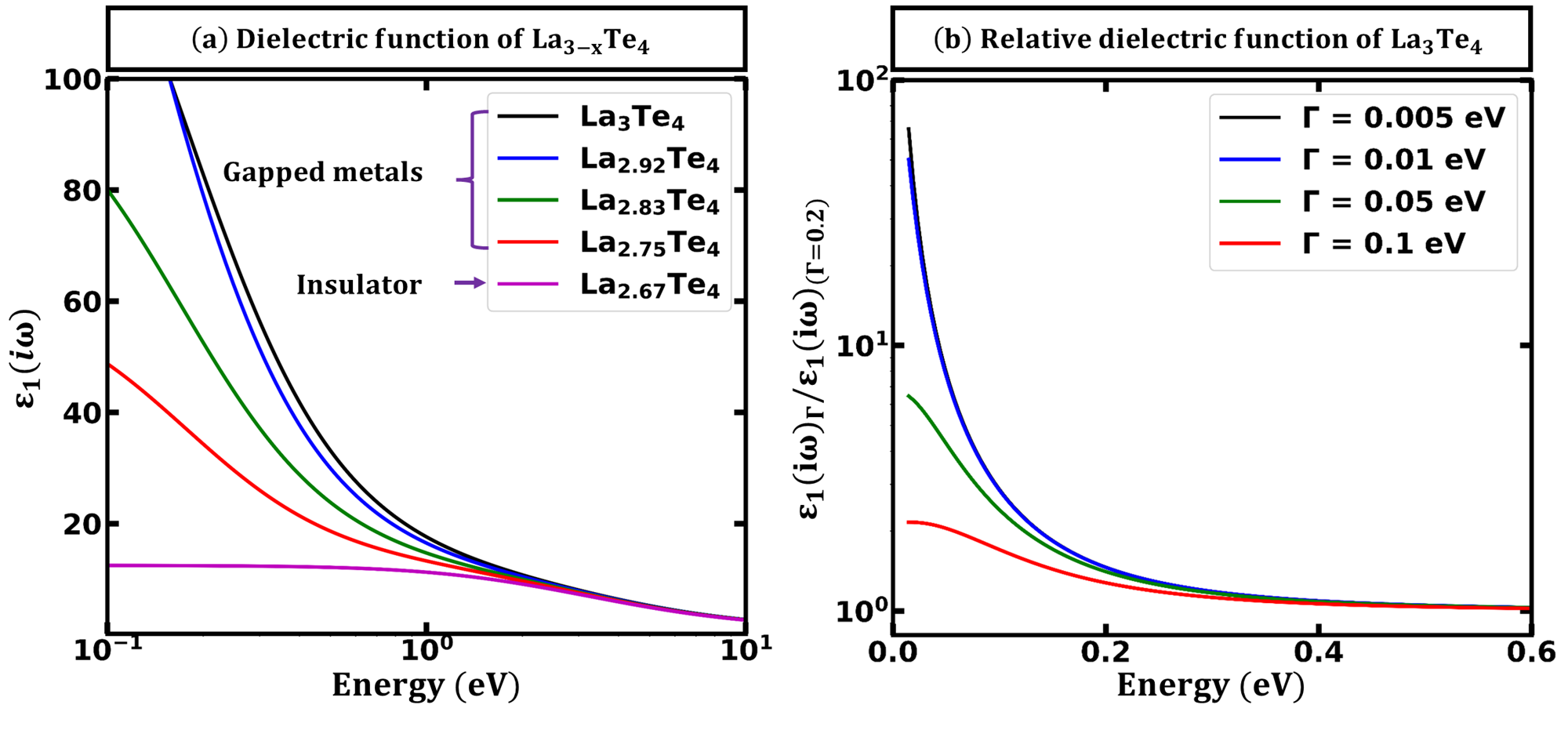}
  \caption{\label{DielectricFunctions}   (Color online) (a) Dielectric functions for imaginary frequencies  from top to bottom for La$_3$Te$_4$, La$_{2.92}$Te$_4$, La$_{2.83}$Te$_4$, La$_{2.75}$Te$_4$, La$_{2.67}$Te$_4$.  (b) We see in the low-frequency region ratio of the dielectric function for La$_3$Te$_4$ with different $\Gamma$ {(damping coefficient also known as dissipation parameter)} to the corresponding values with La$_3$Te$_4$ with $\Gamma=0.2$\,eV. The results include a superposition of interband transitions and Drude (free electron) contributions, both calculated using DFT.}
\end{figure*}

The quantity related to forces follows from the imaginary ($\varepsilon_i''$) part of the dielectric function: \,
\begin{equation}
\varepsilon_i(i \xi_m)=1+\frac{2}{\pi}\int_0^\infty d\omega \frac{ \omega \varepsilon_i''(\omega)}{\omega^2+\xi_m^2},\,{i=1,2,3}
\end{equation}
{ where the Matsubara frequency is $\xi_m=2 \pi k T m/\hbar$, and the subscript \textit{i} indicates the medium.}
As seen in Fig.\,\ref{DielectricFunctions}\, the curves show strong dependence for the dielectric function on off-stoichiometry for La$_{3-x}$Te$_4$ (to be specific: La$_{3}$Te$_4$ {(with E$_g^{int}$ = 1.22 eV)}, La$_{2.92}$Te$_4$ {(with E$_g^{int}$ = 1.16 $\pm$ 0.05 eV)}, La$_{2.83}$Te$_4$ {(with E$_g^{int}$ = 1.16 $\pm$ 0.05 eV)}, La$_{2.75}$Te$_4$ {(with E$_g^{int}$ = 1.16 $\pm$ 0.05 eV)}, La$_{2.67}$Te$_4$ {(with E$_g^{int}$ = 1.13 eV)}) going from a metallic to insulator behavior.  In the Fig.\,\ref{DielectricFunctions}\,b we see for low-frequencies the dependence for the dielectric function of La$_3$Te$_4$ on the $\Gamma$ parameter for La$_3$Te$_4$. { In particular, we present the ratio of the dielectric function for La$_3$Te$_4$ with different $\Gamma$ to the corresponding values with La$_3$Te$_4$ with $\Gamma$ = 0.2 eV. This highlights that it is only at low frequencies the $\Gamma$ parameter influences the dielectric function.} To use the calculated dielectric functions for Casimir-Lifshitz interaction, we also develop the  parametrization of the average dielectric function   (Table 1) with 14-mode oscillator model \cite{Malyi_etal_PCCP_VolDepDielAmorphousSiO2_2016}: 

\begin{equation}
\varepsilon(i \xi)=1+\sum_j \frac{C_{j}}{1+ (\xi/\omega_j)^2},
        \label{ParameteriseddielEq}
\end{equation}
where $\omega_j$ are characteristic frequencies and $c_j$ are proportional to the oscillator strengths. 

To describe the Casimir interaction in systems amenable to experimental measurement, we additionally compute the dielectric properties of gold. These calculations are performed using PBEsol~\cite{R1Perdew2008}  XC functional combined with an effective Hubbard correction of $U=3.0$\, eV on the $d$-orbitals according to Dudarev et al.~\cite{R5DudarevBottonSavrasovHumphreysSutton1998}. Here, the calculations are performed for 1 atom per primitive cell, the k-space summation involved an $80 \times 80 \times 80$ k-mesh and a Gaussian smearing of 0.05\,eV. The resulting permittivity for gold agrees with the experimental data presented in Ref.~\cite{R3aaZoricZachKasemoLanghammer2011}. The plasma frequency was calculated to 9.6\,eV  and we used that with $\Gamma=0.05$\, eV when calculating the Drude contribution.

\subsection{Theory of the Casimir Force}

Let $F(L)$ be the free energy per unit surface area between planar surfaces, and $f(L)$ be the force between a gapped metal (medium 1) and a  sphere with radius $R$ (medium 3).  For $L<<R$ the force between a sphere and a planar surface can be deduced from the free energy between two planar surfaces using the so-called proximity force { approximation}\,\cite{Ser2018}. {This approximation is used to estimate the force between a sphere and a planar surface based on the interaction-free energy between two half-planes.\,\cite{Ser2018}} The sphere may be gold, a gapped transparent conductor, or an air bubble.  The intervening medium 2 may be  a diluted gas with $\varepsilon_2(i \xi_m)\sim 1$, or water with dielectric functions given in the literature\,\cite{JohannesWater2019,luengo2022WaterIce}. The free energy and force can be written as\,\cite{Bordagbook,Ser2018}  
\begin{equation}
F(L)=\frac{f(L)}{2 \pi R} = \frac{k_BT}{2 \pi} {\sum_{m=0}^\infty}{}^\prime \int\limits_0^\infty dq\,q \sum_{\sigma}\ln(1- r_{\sigma}^{21}r_{\sigma}^{23}
 \mathrm e^{-2\kappa_2 L}), \label{LifFreeEnergy}
\end{equation}
where $\sigma=\rm TE,TM$, and the prime in the sum above indicates that the first term ($m$ = 0) has to be weighted by $1/2$. The   Fresnel reflection coefficients between surfaces $i$ and $j$ for the transverse magnetic (TM) and transverse electric (TE) polarizations are given by
\begin{equation}
    r_{\rm TE}^{ij} = \frac{\kappa_i-\kappa_j}{\kappa_i+\kappa_j}\,;  \,\,\,\,\, r_{\rm TM}^{ij} = \frac{\varepsilon_j\kappa_i-\varepsilon_i \kappa_j}{\varepsilon_j \kappa_i+\varepsilon_i \kappa_j} \,. \label{eq:rtTETM}
\end{equation}
Here $\kappa_i= \sqrt{{q}^2+\varepsilon_i\xi_m^2/c^2}$, with $i=1,2,3$ and the { Matsubara frequency is defined above}.
Notably, as the dielectric constant is finite for gapped metals unless we set $\Gamma=0$,  the transverse electric reflection coefficient between non-magnetic gapped metals and dilute gas goes to zero ($r_{\rm TE}^{21}=0 \quad as \quad \xi \to 0$). As a comparison, we also consider what happens when we assume perfect metallic behavior for the transverse electric reflection coefficients in the zero frequency limit\,\cite{Bost2000,PhysRevLett.87.259101,KlimchitskayaMohideenMostepanenko} ($ r_{\rm TE}^{21} r_{\rm TE}^{23}=1 \quad as \quad \xi \to 0$). The alternative limiting behavior leads to the ``perfect metal correction'' (PMC) to the interaction between real metal surfaces (when $L<<R$)\,\cite{Bost2000,PhysRevLett.87.259101,KlimchitskayaMohideenMostepanenko},
\begin{equation}
\Delta F^{\rm PMC}(L)=\frac{\Delta f^{PMC}(L)}{2 \pi R} = \frac{-k_B T \zeta(3)}{16 \pi L^2}, \label{LifFreeEnergyTEn0}
\end{equation}
{ where $\zeta(3)$ is a Riemann zeta function}. This could be the case if the dielectric function for La$_3$Te$_4$ is not finite as $\xi\rightarrow0$, and goes to infinite at least as $1/\xi^2$. The energy correction factor, $\eta(L,T)$, is defined\,\cite{Ser2018} as the ratio of the calculated Casimir force and the corresponding Casimir force between a perfectly conducting sphere with radius $R$ and a perfectly conducting plane (when $L<<R$) at zero temperature\,\cite{casimir1948-1},

\begin{equation}
    F_C(L)=\frac{f_C(L)}{2 \pi R}\cong\frac{-\pi^2 \hbar c}{720 L^3},
    \label{CasForce0K}
\end{equation}

\begin{equation}
     \eta(L,T)=\frac{f(L)}{f_C(L)}=\frac{F(L)}{F_C(L)}\label{RatioCasForce0K}.
\end{equation}
We will match our results against an ideal case (relevant to any metal surfaces behaving as ideal plasmas\,\cite{Bordagbook}).
For two metallic surfaces interacting across air, it is thus relevant for comparison to include the ``perfect metal correction'' by adding Eq.\,(\ref{LifFreeEnergyTEn0}) and Eq.\,(\ref{LifFreeEnergy}) and divide the sum with Eq.\,(\ref{CasForce0K}).  The corrected ratio is,
\begin{equation}
    \eta^*=\frac{f(L)+\Delta f^{PMC}(L)}{f_C(L)}=\frac{F(L)+\Delta F^{PMC}(L)}{F_C(L)}.
    \label{CasForce0KCorrected}
\end{equation}

\subsection{ A note on the effect of dissipation}

{{ The dissipation is included in our paper via a damping term $\Gamma$ in the Drude dispersion relation, thus not  from microscopic models. To see this from a broader perspective, it may be worthwhile to start from the general case where the temperature dependence of $\Gamma$ is also taken into account. We have then
\begin{equation}
\varepsilon(i\xi,T)= 1+ \frac{\omega_{pl}^2}{\xi[\xi +\Gamma(T)]}, \label{1x}
\end{equation}
where $\xi$ is the imaginary frequency. If one at first ignores impurities, one may here make use of the Bloch-Gr{\"u}neisen formula \cite{gray72} for the temperature variation of the electrical resistivity $\rho$, the latter being proportional to $\Gamma$.  This formula implies, among other things, that $\varepsilon (i\xi,T)$ is actually higher when $T$ is low than at room temperature if the frequencies are lower than about $10^{14}~$rad/s, i.e., of the same order as the first Matsubara frequency at room temperature (cf. the discussion on these points by H{\o}ye et al.~\cite{hoye03,brevik05}). Now, from numerical estimates, it turns out that the influence of the temperature on $\Gamma$ is modest. A more important factor in this context is the presence of impurities in the metal. The existence of these makes the resistivity $\rho$, and thus also $\Gamma$, constant at low temperatures and low frequencies.  Consequently, we can put $\varepsilon(i\xi, T) \rightarrow \varepsilon(i\xi) \propto 1/\xi$ when $\xi \rightarrow 0$. Now one must recognize that for practical purposes it is not $\varepsilon$ itself that is the central quantity, but rather the combination
\begin{equation}
 \xi^2[\varepsilon(i\xi)-1], \label{2x}
 \end{equation}
which has to go  to zero as $\xi \rightarrow 0$. This relationship is satisfied by the simple Drude ansatz for the dispersion equation, and the Casimir theory for metals becomes consistent. The expression (\ref{2x}) implies that the contribution to the Casimir force from the TE zero mode vanishes, and also that the Nernst theorem becomes satisfied (i.e., that the free energy depicted as a function of $T$ has a zero slope as $T\rightarrow 0$).  Cf. again the mentioned references by H{\o}ye et al.~\cite{hoye03,brevik05}.
}}

\section{Results}
\subsection{A pair of identical La$_{3-x}$Te$_4$ surfaces}

\begin{figure*}
  \centering
  \includegraphics[width=0.8\textwidth,height=14cm]{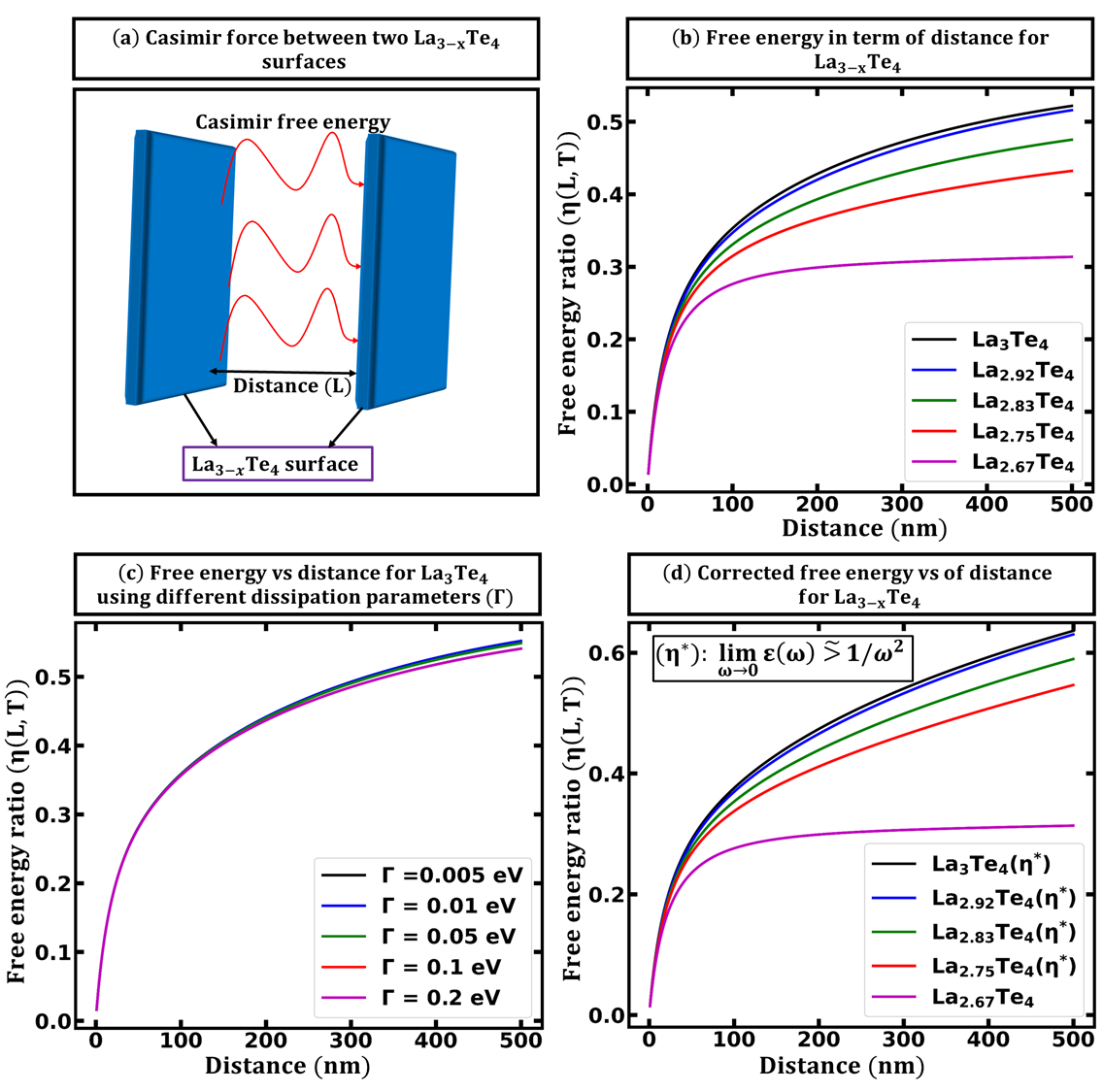}
  \caption{\label{EnergyCorrectionGappedMetals} (Color online) (a) Schematic scheme over the set-up. (b)
  Energy correction factor ($\eta(L,T)$) between equal gapped metal surfaces across air at T=300\,K. (c) $\eta(L,T)$ for two equal La$_{3-x}$Te$_4$ varying the $\Gamma$. (d) Same as in (b) but adding the ``perfect metal correction'' to the free energy for the metallic systems. {  The Drude damping parameter is here introduced on a phenomenological level. The influence from dissipation does not significantly change the strength of the interaction. In Fig. 3d, the reported change  is however different, since it arises from the "perfect metal correction" (the meaning of the zero frequency limit given is explained in Fig.\,\ref{Results_CasimirRatioGoldGapped}).}   }
\end{figure*}

To demonstrate the effect of off-stoichiometry on the Casimir-Lifshitz force, we first consider the interaction between gapped metal planar surfaces interacting across the air as a function of interplane distance (Fig.~\ref{EnergyCorrectionGappedMetals}a). At short separations, that is, when the finite velocity of light can be approximated as infinite, the product of reflection coefficients goes as
\begin{equation}
    \frac{(\varepsilon_1-\varepsilon_2)^2}{(\varepsilon_1+\varepsilon_2)^2}.
    \label{ratio_reflection}
\end{equation}
From this part of the full expression, two things can be deduced: (1) the force between identical surfaces is attractive, and (2) the closer the ratio in Eq.~\eqref{ratio_reflection} is to 1 (i.e. the more metallic), the stronger the attraction.

The Casimir-Lifshitz free energy between gapped metal surfaces at finite temperatures is predicted to deviate strongly from the T=\,0\,K ideal metal Casimir interaction. (Note that the long-range thermal asymptote decays slower than the zero-temperature Casimir asymptote, leading to an increase of the ratio with increasing separations). As can be seen in Fig.~\ref{EnergyCorrectionGappedMetals}b-d, off-stoichiometry effects for gapped metals can be used as an effective knob to induce 10-40\,\% changes in the magnitude of the interaction. Notably, in the separation range where forces are typically measured, the effects are of the same magnitude as any potential corrections from ``the ideal metal'' (plasma model) plate zero frequency transverse electric contribution. The force in Fig.~\ref{EnergyCorrectionGappedMetals}b increases in magnitude as the surfaces become more metallic (up to a limit where the product of the reflection coefficient in a specific frequency range is close to one). 
The results shown in Fig.~\ref{EnergyCorrectionGappedMetals}d is the corresponding case when the ideal metal or "plasma model" approximation\,\cite{Bordagbook} is used for the zero frequency term.

Another important observation is that interaction between different metallic ${\rm La}_{3-x}{\rm Te}_4$ surfaces is also different (i.e., dependent on $x$). This behavior can be understood on the electronic structure theory level as increasing the off-stoichiometry results in a change of free carrier concentration (each La vacancy removes three electrons from the principal conduction band), reducing the free carrier contribution to $\varepsilon(i\xi)$. Hence, even though multiple La$_{3-x}$Te$_4$ surfaces are metallic, they substantially differ in dielectric functions simply due to differences in free carrier concentration.

As noted above, the dielectric properties of materials are defined by the superposition of different contributions. While one can explicitly calculate the band-to-band transition, the calculation of free carrier contribution to dielectric function relies on the Drude model. We described in Fig.~\ref{DielectricFunctions} the effect on the low-frequency tail of the dielectric function from changes in the choice of damping coefficient in the Drude model ($\Gamma$). The static dielectric constant rises to higher values with a reduced $\Gamma$.
However, in Fig.~\ref{EnergyCorrectionGappedMetals}c, we show how slight  the energy correction factor is on the choice of damping coefficient  for the specific case of a pair of La$_{3}$Te$_4$ surfaces. Since the gapped metal surfaces have large but finite dielectric constants, one would expect the ``perfect metal correction'' to be purely hypothetical. However, future optical and force measurements will enable us to distinguish better how to accurately model the dielectric functions at extremely low frequencies.

\subsection{Gold sphere in air near La$_{3-x}$Te$_4$ surfaces}

As is well  known, all  ordinary metals have a finite static conductivity. The Drude model describes the optical and dielectric properties quite well for small frequencies. The dielectric function in the Drude model is  
\begin{equation}
\varepsilon \left( \omega  \right) = 1 + \frac{{i\sigma \left( \omega  \right)}}{\omega } = 1 - \frac{{\omega _{pl}^2}}{{\omega \left( {\omega  + i\Gamma } \right)}}.
\end{equation}
Setting the damping parameter $\Gamma$ zero, the plasma model is obtained. The damping parameter has a real physical origin, and is the result of scattering of the carriers against lattice imperfections. At finite temperature, processes with phonons emitted or absorbed further contribute.  Bostr{\"o}m and Sernelius\,\cite{Bost2000,PhysRevLett.87.259101} found that the damping parameter has a dramatic effect on the Casimir interaction between gold surfaces at separations where the finite temperature is important. The plasma model actually predicts a result that coincides with that of the classic Casimir gedanken experiment between two perfectly reflecting half-spaces, while the Drude model predicts that this result is reduced by a factor of two. In the limit of large $L\rightarrow\infty$ (while still demanding $\frac{L}{R}<<1$), the ratio for two interacting metallic surfaces would, within the PMC, become asymptotically,
\begin{equation}
\eta^*\sim \frac{90 L k_B T \zeta(3)}{\pi^3 \hbar c}, \label{LimRatioTETM}
\end{equation}
while $\eta$ in the same limit has half this magnitude. For all cases considered, the ratios go (for large $L$) $\propto  L k_B T/(\hbar c)$. This means, as is well known, that the ratio increases linearly with separation for large $L$. Some experiments favor the Drude model\,\cite{Lamo1997,Lamoreaux_2005,SushNP} while most appear to favor the plasma model\,\cite{Mohid,RevModPhys.81.1827,Bordagbook,KlimchitskayaMohideenMostepanenko}. The materials we consider are evaluated by density functional theory to have large, but finite, values for the zero frequency permittivity (but should go to infinity if $\Gamma\rightarrow 0$ ). They should hence have a free energy behavior not observed experimentally by, for example, Mohideen and collaborators  for metal surfaces\,\cite{Mohid,RevModPhys.81.1827,Bordagbook,KlimchitskayaMohideenMostepanenko}. Rather it is expected to be closer to the experimental observations by Lamoreaux and his collaborators\,\cite{Lamo1997,Lamoreaux_2005,SushNP} (and even more to the case of doped silicon surfaces). If future experiments would prove this to be wrong, further research must be prompted into how to model the low-frequency part of the dielectric functions for gapped metals within density functional theory.

\begin{figure*}
  \centering
  \includegraphics[width=0.8\textwidth,height=7cm]{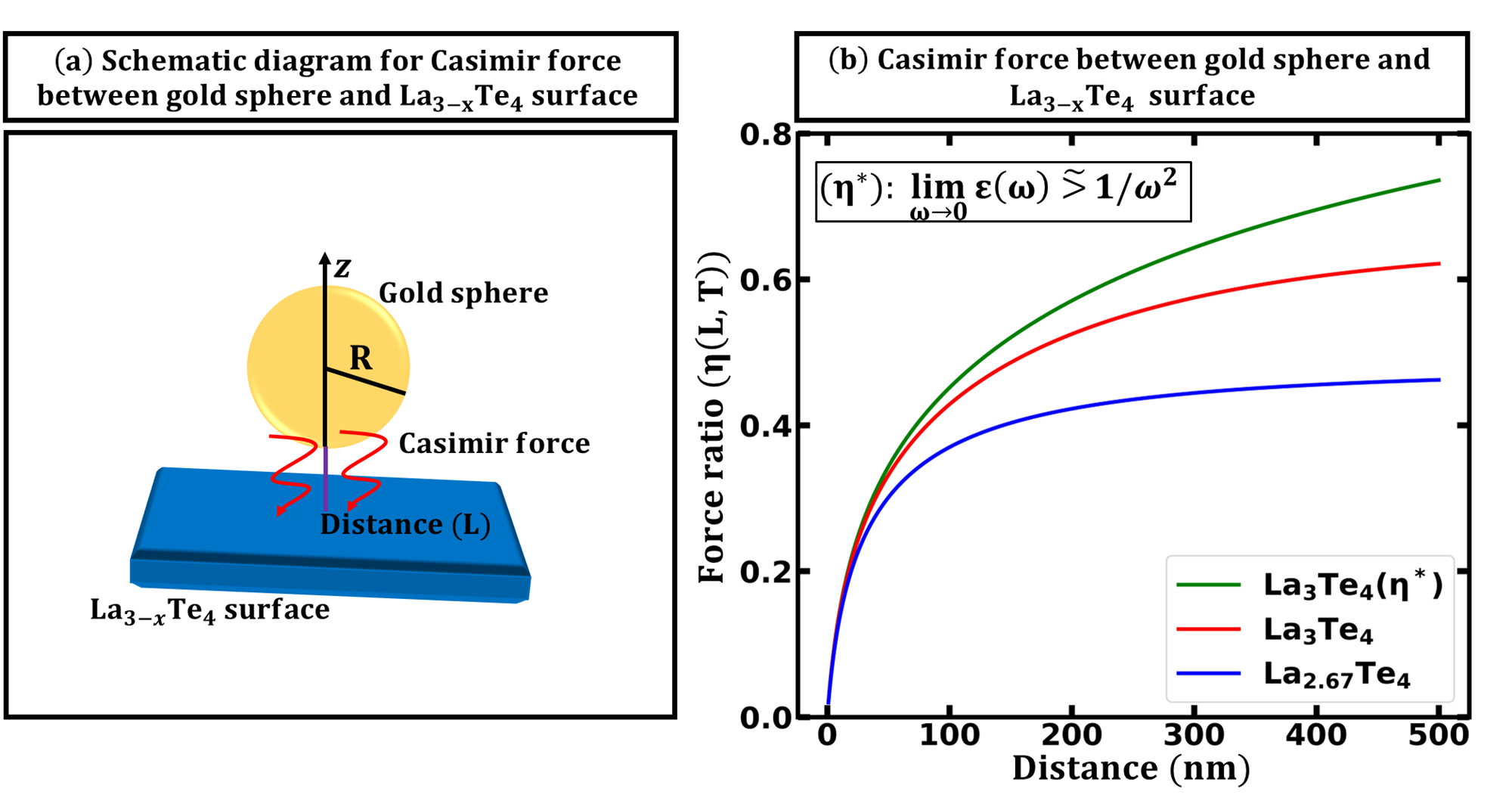}
  \caption{\label{Results_CasimirRatioGoldGapped} (Color online) Schematic set up and the $\eta(L,T)=f(L)/f_C$ at T=300\,K for a gold sphere interacting with (i) La$_3$Te$_4$, (ii) La$_3$Te$_4$ with PMC, i.e. $\xi=0$ transverse electric "Perfect Metal Correction", or (iii) La$_{2.67}$Te$_4$ surface.  { The limiting value ($\varepsilon\rightarrow 1/\omega^2$) refers to the slowest possible decay in the low-frequency limit (i.e. not the actual zero frequency limit) in which the zero frequency TE mode would be expected to contribute.}}
\end{figure*}

Typically, experimental measurements of Casimir forces are conducted using atomic force microscopy (AFM), where a specialized sphere (e.g., Au) is attached to the AFM tip. Because of this, herein, we also consider another proposed experimental setup shown in Fig.~\ref{Results_CasimirRatioGoldGapped}, where we consider Au  sphere attached to an atomic force microscope tip, interacting with a planar surface of different La$_{3-x}$Te$_4$ surfaces at different separation. 
 We present our result in terms of the ratio between the calculated force and the corresponding hypothetical force between a perfectly conducting plane and a perfectly  conducting sphere (in the limit of zero temperature).  Very similar systems, but with different material combinations, have been carefully studied by some of the world's best groups in force measurements. Notably, as we clearly demonstrate, both off-stoichiometry and ``perfect metal correction'' effects are large enough to be within the { few percentage measurement accuracy claimed in the 0.1\,$\mu\,m$ range in several experimental labs (for instance by Mohideen and co-workers)}\,\cite{Lamo1997,HarrisPhysRevA.62.052109,DeccaPhysRevLett.91.050402,PhysRevB.98.201408,SushNP,KlimchitskayaMohideenMostepanenko,SushNP,Mohid, XuGaoShenJacobLiNatComun2022}.

\subsection{Air bubble in liquid water near La$_{3-x}$Te$_4$ surfaces}

One interesting prediction from  theory and force measurements is that these forces can be repulsive\,\cite{Dzya,Rich71,Rich73,MILLING1996460,LEE2002,Feiler2008,Munday2009,Zwol2010,TaborPRL2011}, and even change sign\,\cite{Elbaum,DouPhysRevB.89.201407}.
When reflecting on the expression for the Casimir-Lifshitz interaction, it is clear that repulsion occurs whenever $\varepsilon_{1}>\varepsilon_2>\varepsilon_3$ for a broad range of finite Matsubara frequencies. This was well known to Dzyaloshinskii,   Lifshitz  and   Pitaevskii\,\cite{Dzya}. A remarkable point discussed by Elbaum and Schick\,\cite{Elbaum} is that the dispersion forces { (including\,\cite{Ser2018}, e.g., van der Waals, Casimir-Polder, Lifshitz and Casimir interactions)} can change sign when for a range of ``high'' frequencies (small separations) $\varepsilon_{1}>\varepsilon_2>\varepsilon_3$, while for low frequencies (large separations) $\varepsilon_2>\varepsilon_{1}>\varepsilon_3$. The origin of this effect is the relevant reflection coefficients combined with the exponential factor $e^{-2 \sqrt{{q}^2+\varepsilon_2\xi_m^2/c^2} L}$. At very large separations (the factor $\sqrt{\varepsilon_2} L\xi_m/(q c)$ should be small or of the order unity to result in a significant contribution to the interaction), the finite velocity of light means only the zero frequency term contributes. If
$\varepsilon_2(0)>\varepsilon_{1}(0)>\varepsilon_3(0)$, then long-range attraction follows.

\begin{figure*}
  \centering
  \includegraphics[width=0.8\textwidth,height=7cm]{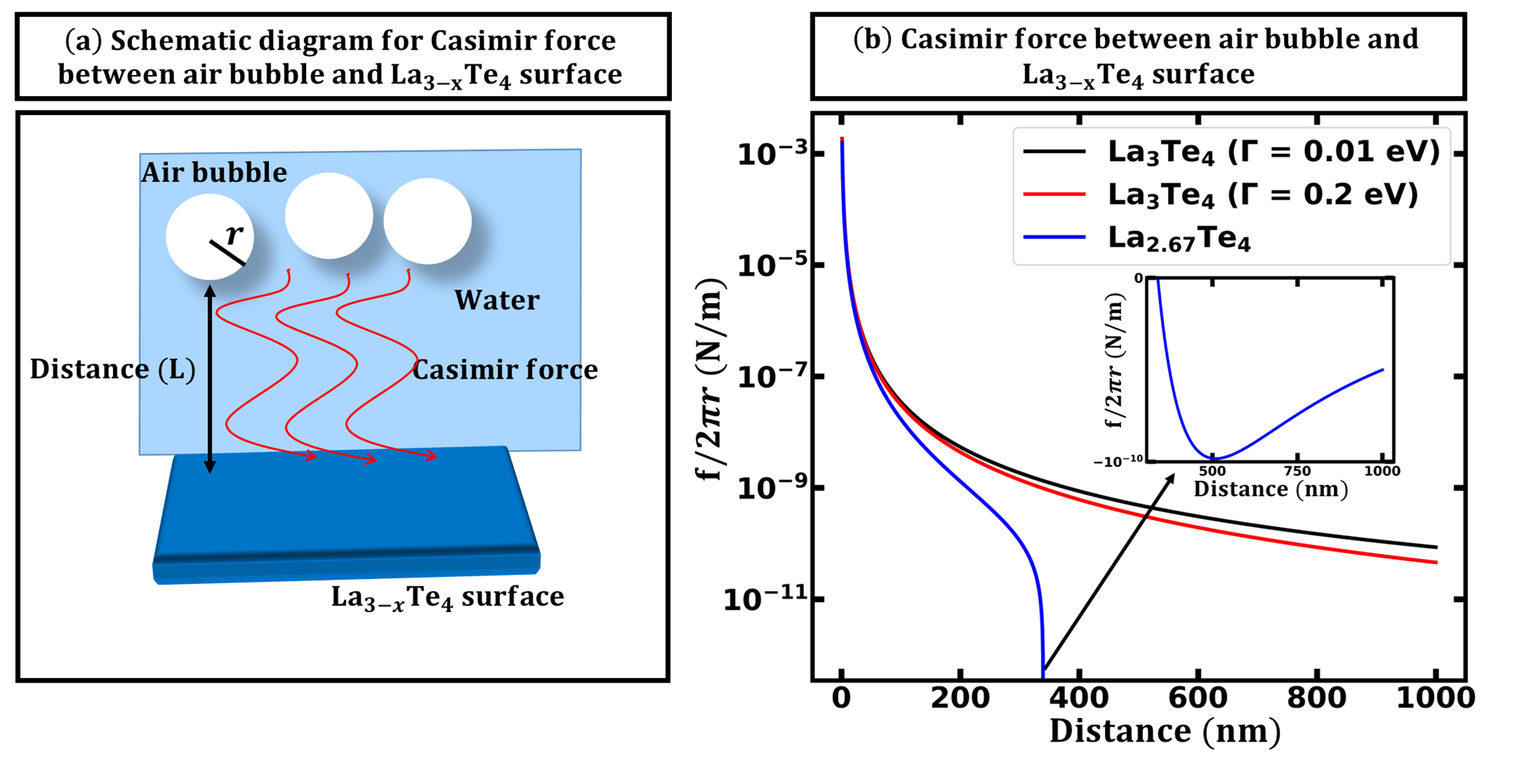}
  \caption{\label{Results_CasimirRatioAirbubbleWaterGapped} (Color online) Casimir force at T=273.16\,K between an air bubble in cold water\,\cite{luengo2022WaterIce} near  (i) La$_3$Te$_4$ with $\Gamma=0.2$\,eV, (ii) La$_3$Te$_4$ with $\Gamma=0.01$\,eV, and (iii) La$_{2.67}$Te$_4$. { The sizes of the bubbles vary depending on the system involved from nanobubbles, micron-sized bubbles to milli bubbles.} }
\end{figure*}

We demonstrate in Fig.~\ref{Results_CasimirRatioAirbubbleWaterGapped} that the Casimir-Lifshitz force between an air bubble in water near a gapped metal surface can, via control of the off-stoichiometry and separation, at large separations change from repulsion to attraction. We propose that further studies on how to tune more effectively the transition from repulsion to attraction-based trapping of gas bubbles in liquids must include other effects including hydration, dissolved gases, surface charges, and ion-specific double layer forces within the DLVO theory and beyond\,\cite{NinhamLoNostro2010book,KunzBook2009}.

\section{Conclusions}

As dimensions and distances relevant for nanomachines go down in size, it becomes increasingly important to control the sign and magnitude of short-range van der Waals/Casimir-Lifshitz interactions\,\cite{doi:10.1126/science.1057984,DelRioBoereedyClewsDunnmicromachined2005,Munday2009} and torques\,\cite{SomersGarrettPalmMunday_CasimirTorque,ThiyamPhysRevLett.120.131601}. 
{ We observe that in the literature one finds that similar analyses have been performed for conductive oxides, phase-changing materials, chiral materials, magnetic material, Weyl semimetals, graphene, topological insulators, and many other systems. We refer the readers to the review by Woods and collaborators\,\cite{WoodsRevModPhys.88.045003}.}
Herein, we investigate the effects of off-stoichiometry on the Casimir-Lifshitz force between La$_{3-x}$Te$_4$ surfaces in various experimental setups. The results show that off-stoichiometry affects significantly impact the interaction between different La$_{3-x}$Te$_4$ surfaces. In particular, we propose that these forces can be influenced to more than 10-40\% by manipulating the off-stoichiometry of gapped metal and tuning it from metallic to insulating behavior (this phenomenon is due to the fact that  La$_3$Te$_4$ gapped metal possesses a distinctive electronic structure resulting in the reduction of free carrier concentration by the formation of La vacancies).  The study of this new material category indicates a  roadmap for how to enhance or reduce Casimir-Lifshitz interactions. As we show, for gas bubbles in liquid water near a gapped metal, tuning off-stoichiometry effects can even change the sign of the long-range part of the Casimir-Lifshitz interactions. Our main conclusion is that off-stoichiometry in gapped metals can be used as a knob to tune long-range interactions. In a longer perspective,  the use  of controlled quantum switches for fluid systems, together with an exploitation  of  off-stoichiometry effects, may be a promising way  to investigate the Drude/plasma controversy.

\begin{acknowledgments}
The authors thank the "ENSEMBLE3 - Centre of Excellence for nanophotonics, advanced materials and novel crystal growth-based technologies" project (GA No. MAB/2020/14) carried out within the International Research Agendas programme of the Foundation for Polish Science co-financed by the European Union under the European Regional Development Fund, the European Union's Horizon 2020 research and innovation programme Teaming for Excellence (GA. No. 857543), and European Union's Horizon 2020 research and innovation programme (grant No. 869815), for support of this work.  We gratefully acknowledge Poland's high-performance computing infrastructure PLGrid (HPC Centers: ACK Cyfronet AGH) for providing computer facilities and support within computational grant no. PLG/2022/015458. We also acknowledge access to high-performance computing resources via NAISS, provided by NSC and PDC, as well as NOTUR, provided by Sigma2.
\end{acknowledgments}

\bibliography{StoichCasimir}

\end{document}